\documentclass[a4paper,11pt]{article}
\pdfoutput=1 % if your are submitting a pdflatex (i.e. if you have
\usepackage{jcappub} % for details on the use of the package, please
\title{\boldmath Spectral shape analysis for
electron antineutrino oscillation study
by using $^{8}$Li generator
with $^{252}$Cf source}

\author[a]{Jae Won Shin,}
\author[a,1]{Myung-Ki Cheoun,\note{Corresponding author.}}
\author[b,c]{Toshitaka Kajino }
\author[d]{and Takehito Hayakawa}

\affiliation[a]{Department of Physics,
Soongsil University,\\Seoul 156-743, Korea}
\affiliation[b]{Division of Theoretical Astronomy,
National Astronomical Observatory
of Japan,\\Mitaka, Tokyo 181-8588, Japan}
\affiliation[c]{Department of Physics, Graduate School of Science, University of
Tokyo,\\Hongo, Bunkyo-ku, Tokyo 113-0033, Japan}
\affiliation[d]{National Institute for Quantum and Radiological Science and Technology,\\ 2-4 Shirakata, Tokai, Naka, Ibaraki 319-1106, Japan}

\emailAdd{shine8199@skku.edu}
\emailAdd{cheoun@ssu.ac.kr}
\emailAdd{kajino@nao.ac.jp}
\emailAdd{hayakawa.takehito@qst.go.jp}

\date{\today}

\abstract{
Existence of hypothetical fourth neutrino,
so-called sterile neutrino,
is one of open issues in the particle and neutrino physics.
This fourth neutrino is a candidate for explaining
some anomalies reported in
LSND, MiniBoone, reactor experiments, and gallium experiments.
To search for the existence of the sterile neutrino,
we report detailed analysis of a feasible experiment for short baseline
electron antineutrino (${\bar{\nu}}_{e}$)
disappearance study, in which a ${\bar{\nu}}_{e}$ source
from $^{8}$Li generator is considered under non-accelerator system.
For $^{8}$Li production,
we suggest to use $^{252}$Cf source as an intense neutron emitter,
by which one can produce $^{8}$Li isotope through $^{7}$Li(n,$\gamma$)$^{8}$Li reaction, effectively.
Using the $^{8}$Li generator,
one does not need any accelerator
or reactor facilities
because the generator
can be placed on any present and/or planned neutrino detectors
as closely as possible.
For the effect of the possible sterile neutrinos,
we estimate expected neutrino flux and event rates from the neutrino source scheme,
and show neutrino disappearance features and possible reaction rate changes
by the sterile neutrino using the spectral shape analysis.
}


\begin{document}
\maketitle
\flushbottom

\section{Introduction}

Since the first observation of neutrino oscillation phenomena
was done in Homestake Experiments \cite{oscil_exp_1},
many experiments (e.g. KamioKanDe \cite{Kamio_1} and SNO \cite{SNO_1} facilities)
had confirmed the neutrino oscillation phenomena
and showed that the oscillation mechanism
resolves the long-standing discrepancies among measured and/or theoretical solar neutrino flux,
so-called solar neutrino problem.

There are `somewhat similar' discrepancies or anomalies, similar to the solar neutrino problem, that were observed in LSND \cite{LSND_1}, MiniBoone \cite{MiniB_1},
reactor antineutrino experiments \cite{reacAnt}
and gallium experiments \cite{GaAnomaly}
despite the knowledge that three flavor neutrinos oscillate.
These observations naturally lead to the possible existence of hypothetical fourth neutrino,
so-called sterile neutrinos (${\nu}_{s}$),
which may mix with standard three active neutrinos
and do not interact with other particles.
For the ${\nu}_{s}$ search, many interesting studies with several neutrino sources
such as $\sim$100 kCi of $^{144}$Ce-$^{144}$Pr antineutrino generators \cite{Ce144s2, Ce144s3}
and electron antineutrinos (${\bar{\nu}}_{e}$) from $^{8}$Li
by using an accelerator-based IsoDAR concept \cite{annu8Li1, sterileNu1}, etc.
have been proposed.

In the previous work,
for the ${\nu}_{s}$ search, we proposed a fissionable isotope of $^{252}$Cf as a radioactive isotope-based
${\bar{\nu}}_{e}$ production scheme \cite{NS252Cf}.
With $^{252}$Cf radioactive isotope, 99.99\% enriched $^{7}$Li and graphite,
$^{8}$Li isotopes as a ${\bar{\nu}}_{e}$ source can be produced effectively.
$^{252}$Cf emits neutrons with an average energy of approximately 2 MeV,
and the generated neutron can
produce $^{8}$Li isotopes by the neutron capture reaction to $^{7}$Li.
Furthermore, $^{8}$Li generator
can be placed on
existing and/or planned
any neutrino detectors
such as Borexino, JUNO, KamLAND, LENA and SNO+, etc.
because one does not need any
accelerator or reactor systems.

In this work,
we study the experimental method by using spectral shape analysis
with the neutrino source from the $^{8}$Li generator
based on a $^{252}$Cf neutron source.
Spectral shapes of the measured neutrinos
can be analyzed independently of absolute flux values and
give valuable chances to study the existence of fourth neutrinos.
To decipher the effect of possible sterile neutrinos
from the shape analysis exploited in Daya Bay \cite{ReactBump_Day}, Double Chooz \cite{ReactBump_DOU1},
and RENO \cite{ReactBump_RENO1, ReactBump_RENO2}
experiments,
we calculate expected neutrino flux and event rates,
and discuss neutrino disappearance features and possible reaction rate changes
by the sterile neutrino.
To simulate the non-accelerator
$^{8}$Li generator,
we use particle transport Monte Carlo code, GEANT4 \cite{g4n1, g4n2}.

\section{Methods
\label{meth-sec}}

\subsection{Electron antineutrino sources from $^{8}$Li generator under non-accelerator system
\label{meth-sec_source}}

As an intense neutron emitter to generate $^{8}$Li,
we consider a $^{252}$Cf isotope with a half-life (T$_{1/2}$) of 2.64 yr,
which emits neutrons through spontaneous fission process.
The following energy distribution of the neutron
from $^{252}$Cf known as Watt fission spectrum \cite{watt_1, watt_2, x5},
$f(E) = {\rm exp}(-\frac{E}{1.025}){\rm sinh}(2.926E)^{1/2}$ is adopted, where {\em E} is the neutron energy in MeV,
and the neutron emission rate of
2.34 $\times$ 10$^{12}$ neutrons per second (n/s) for
1 g of $^{252}$Cf
is used in our work.
These neutrons can produce $^{8}$Li isotope through
$^{7}$Li(n,$\gamma$)$^{8}$Li reaction
where the $^{8}$Li becomes electron antineutrino source.

%%%%%%%%%%%%%%%%%%%%%%%%%%%%%%%%%%%%%%%%%%%%%%%%%%%%%%%%%%%%%%%%%%%%%%%%%%%%%
\begin{figure}[tbp] %Fig1
\centering
\epsfig{file=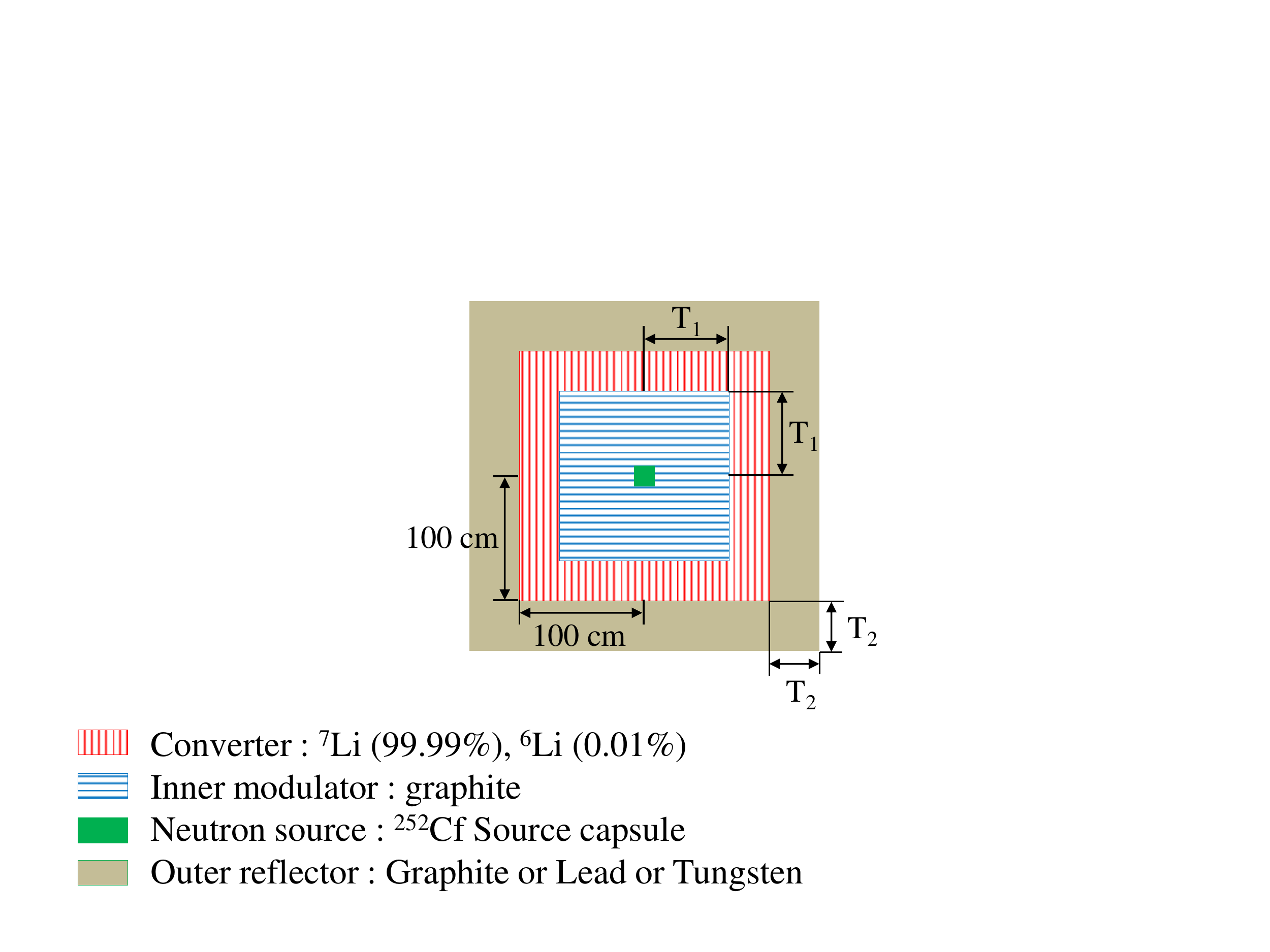, width=5.in}
\caption{(Color online)
A schematic cross section view showing cylindrical $^{8}$Li generator.
}
\label{fig1}
\end{figure}
%%%%%%%%%%%%%%%%%%%%%%%%%%%%%%%%%%%%%%%%%%%%%%%%%%%%%%%%%%%%%%%%%%%%%%%%%%%%%

Figure \ref{fig1} shows the $^{8}$Li generator
comprising 99.99\% enhanced $^{7}$Li convertor,
$^{252}$Cf source surrounded by the graphite modulator, and reflector materials such as C, Pb, W which wrap outside the Li convertor.
The Li convertor has a cylindrical shape of a radius 100 cm
and a length 200 cm, whose shape and size are based on the analysis performed in Ref.~\cite{sterileNu1}.
We place graphite as an neutron modulator
in the $^{7}$Li convertor.
The graphite is usually preferred as a modulator (or reflector)
due to its low absorption cross section
and high elastic scattering cross section for neutron.\footnote{\protect
This concept known as
the Adiabatic Resonance Crossing (ARC)
was first proposed by Nobel Laureate Carlo Rubia \cite{ARC_0}
for the transmutation of long-lived nuclear waste
and the medical radioisotope production \cite{ARC_1, ARC_2, ARC_3}.
}
With the graphite modulator,
we can increase the production yield of $^{8}$Li
even though amounts of the $^{7}$Li are less than those
without graphite.

Heat production of the source can be issued for the experimental setup.
Heat production of approximately 50 kCi of $^{144}$Ce-$^{144}$Pr antineutrino generators
and electron antineutrinos (${\bar{\nu}}_{e}$) from $^{8}$Li
by using an accelerator-based IsoDAR concept are $\sim$370 W and 600 kW,
respectively.
In contrast,
the heat of a $^{252}$Cf source with 1.0 g
is approximately 38.5 W
(decay heat of $\alpha$-decay and spontaneous fission are 18.8 W/g and 19.7 W/g, respectively).
This is much smaller than those from the above generators,
and thus the cooling of 38.7 W is more easy.
Also, in our design,
because the $^{252}$Cf source inside of the graphite and the detector can be located apart,
the cooling of the neutrino source is more easy, if needed.

To calculate production rates of $^{8}$Li
by neutrons from $^{252}$Cf,
the GEANT4 (GEometry ANd Tracking) code \cite{g4n1, g4n2} is used.
For an accurate simulation of neutron interactions,
high precision models (G4HP)
with G4Neutron Data Library (G4NDL) 4.5
are used in the present work,
where the G4HP include cross sections and final states information
for elastic, inelastic scattering,
capture, fission, and isotope production.
The data in G4NDL 4.5 come largely from
the Evaluated Nuclear Data File (ENDF/B-VII) library \cite{endf}.\footnote{\protect
ENDF/B-VII library is developed and maintained by
the Cross Section Evaluation Working Group (CSEWG)
where the data are based on experimental and theoretical data.
}
% beta- decay
$^{8}$Li with $T_{1/2}$ of 0.838 s emits
${\bar{\nu}}_{e}$
through $\beta^{-}$ decay,
${\rm {^{8}Li}} \to {\rm {^{8}Be}} + e^{-} + {\bar{\nu}}_{e}$.
The energy distribution of the
electron anti-neutrinos from $^{8}$Li
is calculated by using
``G4RadioactiveDecay" \cite{G4RDM_0, G4RDM_1} class
based on the Evaluated Nuclear Structure Data File (ENSDF) \cite{G4ensdf}.

\subsection{Proposed experimental setup
\label{sec2_2}}

%%%%%%%%%%%%%%%%%%%%%%%%%%%%%%%%%%%%%%%%%%%%%%%%%%%%
\begin{figure}[tbp]
\begin{center}
\epsfig{file=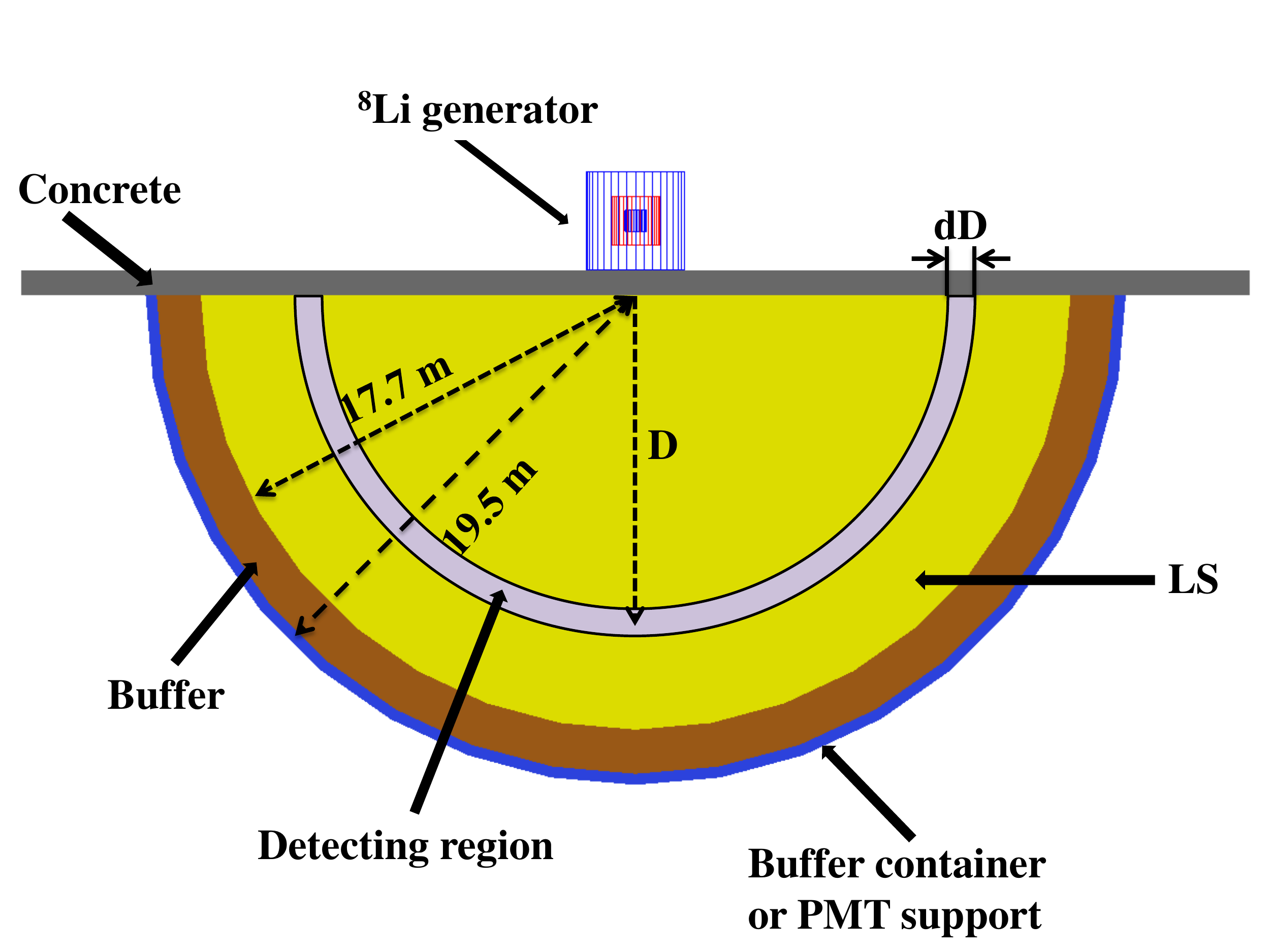, width=3.in}
\end{center}
\caption{(Color online)
Simulation geometry for hemisphere shape detector with $^{8}$Li generator.
}
\label{figG_IBD}
\end{figure}
%%%%%%%%%%%%%%%%%%%%%%%%%%%%%%%%%%%%%%%%%%%%%%%%%
%

To study spectral shape analysis for the ${\nu}_{s}$ existence,
a hemisphere shape liquid scintillator detector
based on the JUNO \cite{JunoR} is considered.
Figure \ref{figG_IBD} shows the simulation geometry
for hemisphere shape detector with the $^{8}$Li generator where
D and dD mean the distance between the bottom of the plate and the center of the detection shell
and the thickness of the shell, respectively.
In this work,
we consider a hemisphere shape of LS detectors with
a radius of 17.7 m
where the target proton number ($n_{p}$)
in the detector is
0.725 $\times$ 10$^{33}$
(approximately 10 kt) \cite{JunoR}.
The expected event rate is obtained within the detecting region
of LS detectors for various D values
with the cylindrical $^{8}$Li generator.

\subsection{Detection of electron antineutrino with liquid-scintillator detector
\label{sec2_3}}

For detection of electron antineutrinos,
an inverse beta decay (IBD) reaction,
${\bar{\nu}}_{e} + p \to e^{+} + n$, is considered.
The IBD reaction gives two distinct signals in electron anti-neutrino detections;
one is the prompt signal due to an annihilation of a positron,
and another is a delayed signal of a 2.2 MeV $\gamma$-ray
via neutron capture.
The event rate ($ER^{IBD}_{{\bar{\nu}}_{e}}$) for IBD
can be written as
\begin{eqnarray}
ER^{IBD}_{{\bar{\nu}}_{e}}
= n_{p} \int^{E_{max}}_{E_{th}} dE_{\bar{\nu}} \Phi_{{\bar{\nu}}_{e}} (E_{\bar{\nu}}) {\rm{P}}_{\nu \bar{\nu}}(E_{\bar{\nu}}) \sigma^{IBD}_{{\bar{\nu}}_{e}}(E_{{\bar{\nu}}_{e}})~,
\label{eq:IBD_rate}
\end{eqnarray}
where $n_{p}$ is the number of target protons within the fiducial volume of detector.
$\Phi_{{\bar{\nu}}_{e}} (E_{\bar{\nu}})$ is the electron-antineutrino flux from $^{8}$Li,
$E_{max}$ is the maximal neutrino energy,
$E_{th}$ is the threshold energy of the reaction,
${\rm{P}}_{\nu \bar{\nu}}(E_{\bar{\nu}})$ is an energy dependent electron-antineutrino survival probability,
and $E_{\bar{\nu}}$ is energy of the incident antineutrino.
The energy dependent cross section of the IBD
in Eq.~(\ref{eq:IBD_rate})
can be expressed by \cite{LENA_ER, ibdCS}
\begin{eqnarray}
\sigma^{IBD}_{{\bar{\nu}}_{e}}(E_{{\bar{\nu}}_{e}})
  \approx p_{e} E_{e}
  E_{{\bar{\nu}}_{e}}^{-0.07056+0.02018{\rm{ln}}E_{{\bar{\nu}}_{e}} - 0.001953 {\rm{ln}}^{3} E_{{\bar{\nu}}_{e}}} \times 10^{-43} [\rm{cm}^{2}],
\label{eq:IBDcs}
\end{eqnarray}
where $p_{e}$, $E_{e}$ and $E_{{\bar{\nu}}_{e}}$ are
the positron momentum, total energy of the positron
and the energy of ${\bar{\nu}}_{e}$ in MeV, respectively.
$E_{e} = E_{{\bar{\nu}}_{e}} - \Delta$ where $\Delta$ is mass difference
between $m_{n}$ and $m_{p}$ ($\Delta = m_{n} - m_{p} \approx 1.293 ~\rm{MeV}$).
This cross section agrees within few per-mille
with the full calculation including
the radiative corrections and the final-state interactions
in IBD.\footnote{\protect
There is another possible ${\bar{\nu}}_{e}$ detection channel, ${\bar{\nu}}_{e}$-e$^{-}$ elastic scattering (ES).
Through the ${\bar{\nu}}_{e}$-e$^{-}$ ES,
antineutrinos can be indirectly measured
by the outgoing scattered electron
which can be identified
by means of the scintillation light
produced in the liquid-scintillator (LS).
But, reaction rates for the ES are much smaller than those for IBD. Therefore, we only consider IBD reaction for the following neutrino disappearance study.
}

\subsection{Tested hypothetical models for the sterile neutrinos
\label{sec2_4}}

By using our ${\bar{\nu}}_{e}$ source,
we can study the possible existence of the fourth neutrino, sterile neutrino.
For three neutrino oscillation model,
we use ${\rm{P}}_{\nu {\bar \nu}}(E_{\bar{\nu}})$ ($\equiv$P$_{3}$)
given by \cite{PeeRef1}
\begin{eqnarray}
%P(\bar{\nu}_{e} {\to} \bar{\nu}_{e})
P_{3}
= 1 - {\rm{sin}}^{2} 2 \theta_{13} S_{23} - c^{4}_{13}{\rm{sin}}^{2} 2 \theta_{12} S_{12},
\label{eq:Pee_1}
\end{eqnarray}
where
$S_{23} = {\rm{sin}}^{2}(\Delta m^{2}_{32} L / 4 E)$ and
$S_{12} = {\rm{sin}}^{2}(\Delta m^{2}_{21} L / 4 E)$.
L and E mean the source to detector distance and
the neutrino energy, respectively.
Neutrino oscillation parameters
in Eq.~(\ref{eq:Pee_1})
are taken from a global fit from Ref.~\cite{nu3active}.

To see the ${\nu}_{s}$ effect,
we also use electron-antineutrino survival probabilities
in the 3+1 and 3+2 models where the survival probabilities
can be written as \cite{sterileNu1}
\begin{eqnarray}
P_{\rm{3+1}}
= 1 - 4|U_{e4}|^{2}(1-|U_{e4}|^{2}){\rm{sin}}^{2}(\Delta m^{2}_{41} L / 4E),
\label{eq:P_31}
\end{eqnarray}
\begin{eqnarray}
P_{\rm{3+2}} =
1 &-& 4[ (1-|U_{e4}|^{2}-|U_{e5}|^{2}) \nonumber \\
&\times& (|U_{e4}|^{2}{\rm{sin}}^{2}(\Delta m^{2}_{41} L / 4E)
+ |U_{e5}|^{2} {\rm{sin}}^{2}(\Delta m^{2}_{51} L / 4E)) \nonumber \\
&+& |U_{e4}|^{2} |U_{e5}|^{2} {\rm{sin}}^{2}(\Delta m^{2}_{54} L / 4E)].
\label{eq:P_32}
\end{eqnarray}
The oscillation parameters for the 3+1 model in Eq.~(\ref{eq:P_31})
and for the 3+2 model in Eq.~(\ref{eq:P_32})
are taken from the best-fit points
from the combined short base lines (SBL) and IceCube data set \cite{s3p1Para}, and
reactor antineutrino data \cite{sterileNu2}, respectively.
%(see Table 1. in Ref.~\cite{sterileNu2}).

\section{Results
\label{results-r1}}

\subsection{$^{8}$Li isotope yield in the generator}

First,
we calculate the production yield of $^{8}$Li
with the $^{252}$Cf source coupled only with the Li convertor.
Yield of 0.0045 $^{8}$Li per neutron ($^{8}$Li/n)
is obtained because neutrons can easily escape from the convertor,
so that they are not effectively captured by $^{7}$Li isotopes.
When graphite material,
the inner modulator in Fig. \ref{fig1},
is placed in the Li convertor,
$^{8}$Li yield can be increased
compared to that without the graphite.
As the graphite thickness increases,
numbers of the collisions
between the neutron and
the carbon nuclei also increase.
Consequently, capture probability of the neutron by $^{7}$Li
can also increase.
If the graphite is too thick, however,
it becomes hard for the scattered neutron
to escape from the graphite.
We found that
yields of $^{8}$Li increase
when T$_{1}$ increases up to 43 cm.
As the T$_{1}$
of the inner modulator
increases more than 43 cm,
yields of $^{8}$Li decrease.
Maximum yield of $^{8}$Li turns out to be 0.1572 $^{8}$Li/n at T$_{1}$ = 43 cm,
whose yield is about 35 times larger than that without graphite.

%%%%%%%%%%%%%%%%%%%%%%%%%%%%%%%%%%%%%%%%%%%%%%%%%
\begin{figure}[tbp]
\begin{center}
\epsfig{file=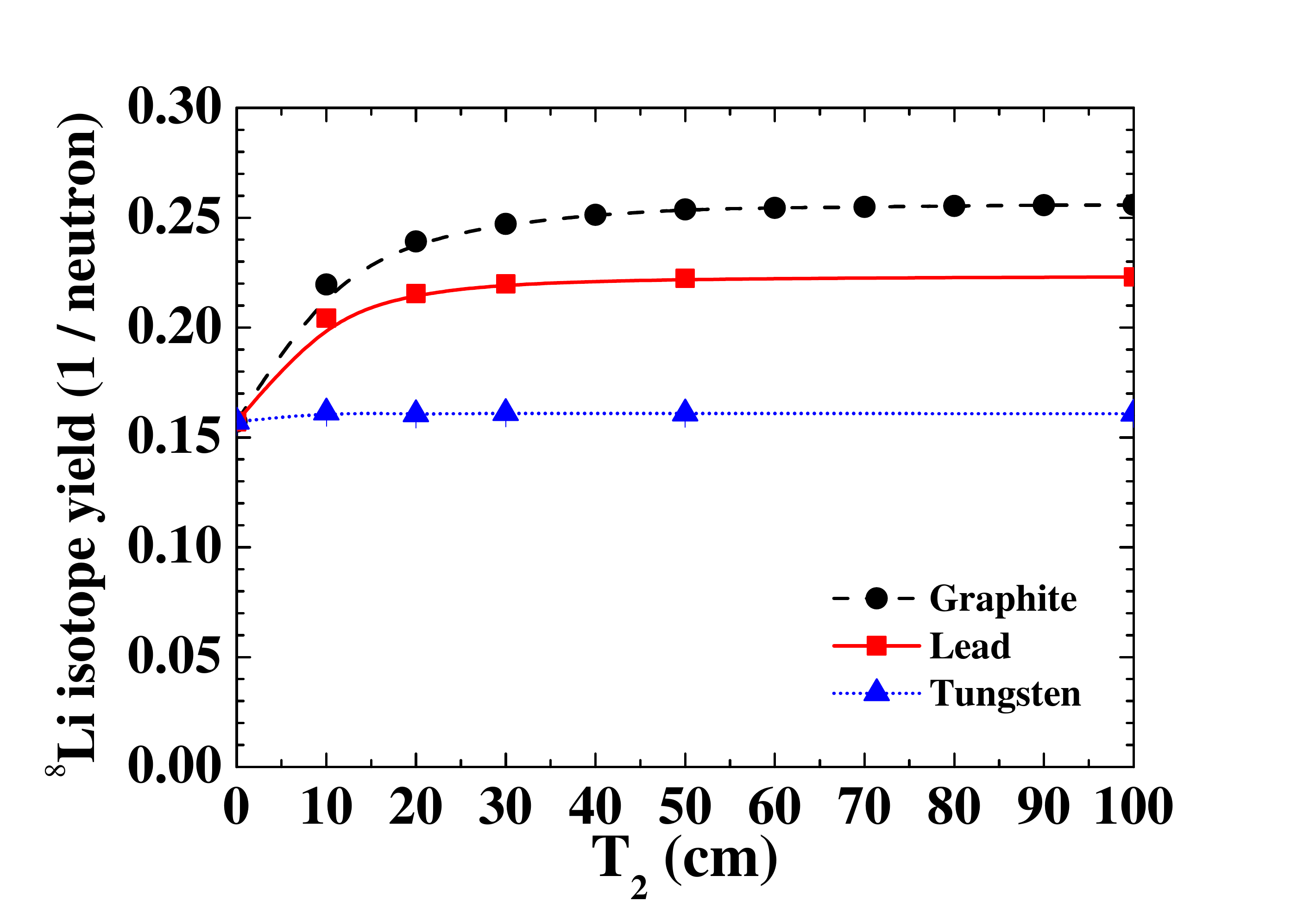, width=4.in}
\end{center}
\caption{(Color online)
$^{8}$Li isotope yields for graphite, lead and tungsten
with respect to T$_{2}$.
T$_{1}$ is set to 43 cm because the value is optimal thickness.
The filled circles, squares and triangles denote results for
graphite, lead and tungsten, respectively.
}
\label{set2gra}
\end{figure}
%%%%%%%%%%%%%%%%%%%%%%%%%%%%%%%%%%%%%%%%%%%%%%%%%%%

To further increase the yields of $^{8}$Li,
outer reflector
with the thickness T$_{2}$ in Fig. \ref{fig1}
is considered to be located out of the Li convertor.
The production yield for $^{8}$Li
with respect to T$_{2}$
is plotted in Fig.~\ref{set2gra}
where the optimal thickness of T$_{1}$ is chosen as 43 cm.
As the reflector materials, graphite, lead and tungsten are considered.
For graphite,
$^{8}$Li yields increase up to 0.256 $^{8}$Li/n
as the T$_{2}$ increases.
With the T$_{2}$ larger than 50 cm,
yields for $^{8}$Li are almost saturated.
Results for lead and tungsten as the T$_{2}$ material
are also shown in Fig. \ref{set2gra}.
The production yields of $^{8}$Li for
lead and tungsten increase up to 0.22 and 0.16 $^{8}$Li/n, respectively,
with the increase of the T$_{2}$.
Yields of $^{8}$Li
for lead and tungsten as the T$_{2}$ material
are smaller than that of graphite
by $\sim$ 16\% and $\sim$ 60\%, respectively.
It is found that yield of 0.256 $^{8}$Li/n is obtained
with both the two graphite materials (modulator and reflector).
Therefore,
we use this value for the setup in this work.

%%%%%%%%%%%%%%%%%%%%%%%%%%%%%%%%%%%%%%%%%%%%%%%%%%%
\begin{figure}[tbp]
\begin{center}
\epsfig{file=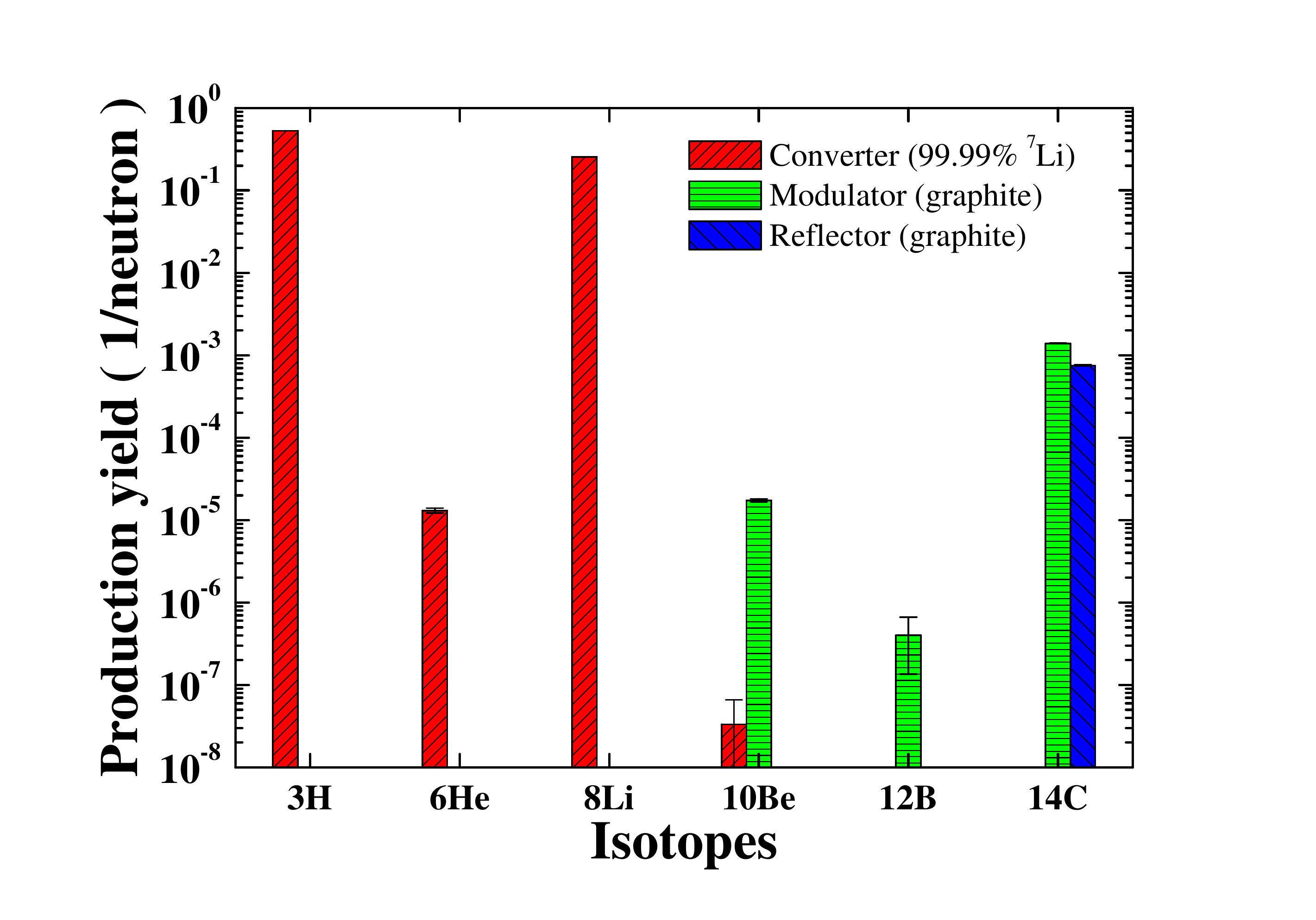, width=4.in}
\end{center}
\caption{(Color online)
Production yields of unstable isotopes in the $^{8}$Li generator.
The modulator and the reflector materials are chosen as graphite, and
T$_{1}$ and T$_{2}$ are 43 cm and 100 cm, respectively.
}
\label{fomu_bench}
\end{figure}
%%%%%%%%%%%%%%%%%%%%%%%%%%%%%%%%%%%%%%%%%%%%%%%%%

%%%%%%%%%%%%%%%%%%%%%%%%%%%%%%%%%%%%%%%%%%%%%%%%%%%
\begin{figure}[tbp]
%\begin{center}
\centering
\includegraphics[scale=0.5]{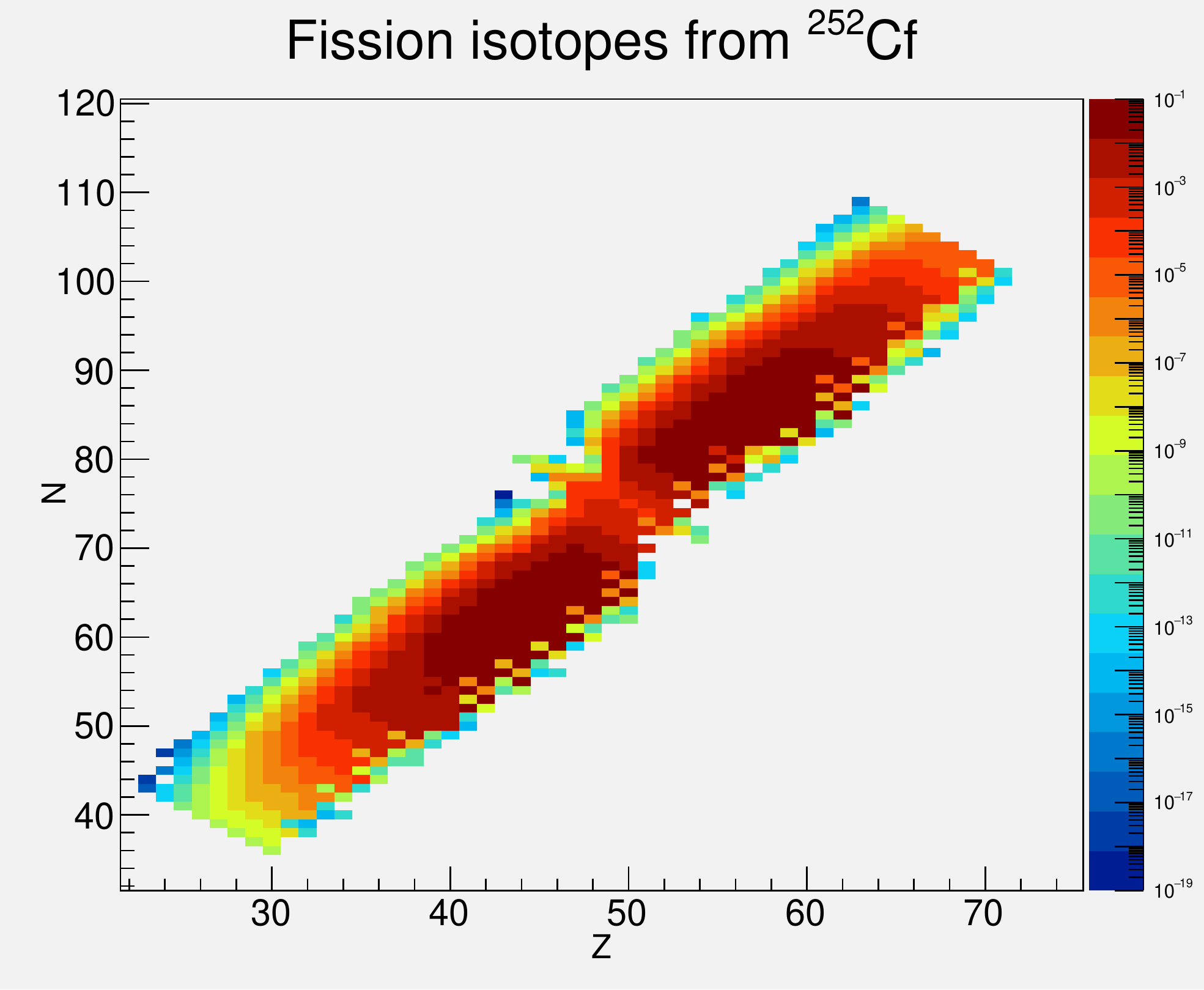}
%\end{center}
\caption{(Color online)
Yields of fission isotopes from $^{252}$Cf.
}
\label{Fy_252Cf}
\end{figure}
%%%%%%%%%%%%%%%%%%%%%%%%%%%%%%%%%%%%%%%%%%%%%%%%%

\subsection{Background consideration
\label{results-bgc}}

The IBD reactions provide two distinct signals in electron anti-neutrino detection;
one is the prompt signal due to an annihilation of a positron,
and another is a delayed signal of a 2.2 MeV $\gamma$
via a neutron capture
that provides almost unambiguous antineutrino event detection.
%Characteristics of the two distinct detections
These coincident features
give an efficient rejection
of other possible backgrounds.

%%%%%%%%%%%%%%%%%%%%%%%%%%%%%%%%%%%%%%%%%%%%%%%%%%%
\begin{figure}[tbp]
%\begin{center}
\centering
\epsfig{file=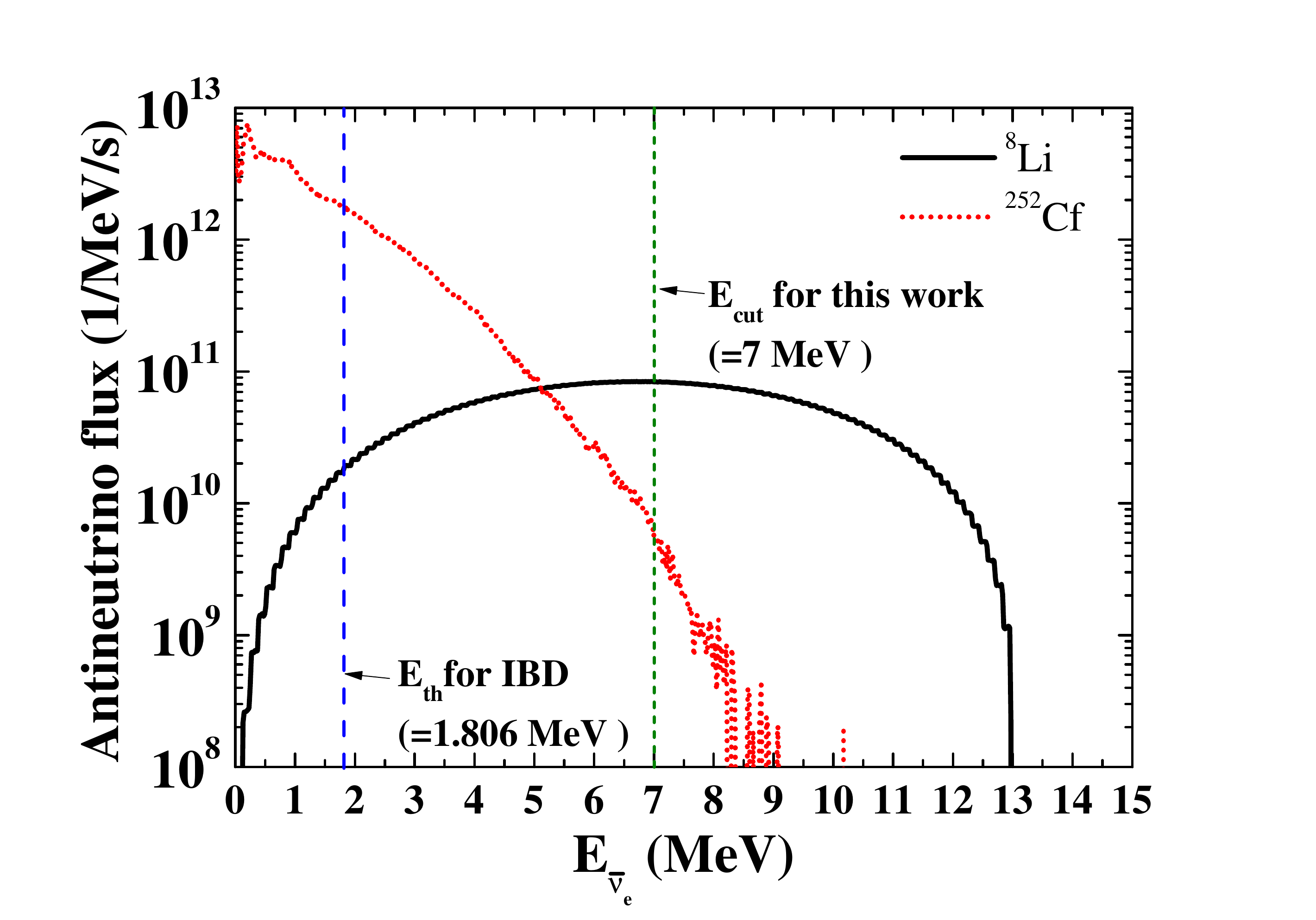, width=4.in}
%\end{center}
\caption{(Color online)
${\bar{\nu}}_{e}$ flux from $^{8}$Li and fission products of $^{252}$Cf.
The black solid lines denote the electron-antineutrino from $^{8}$Li
and the red dotted lines represent those from fission products of $^{252}$Cf.
}
\label{figFISS}
\end{figure}
%%%%%%%%%%%%%%%%%%%%%%%%%%%%%%%%%%%%%%%%%%%%%%%%%

In fact, various unstable isotopes
can be produced in
both Li convertor and graphite modulator and reflectors, and emit antineutrinos. Thus they can affect the neutrino detection.
Production yields of unstable isotopes
for T$_{1}$ = 43 cm and T$_{2}$ = 100 cm
are plotted in Fig.~\ref{fomu_bench}.
Figure \ref{fomu_bench} shows that
$^{3}$H, $^{6}$He and $^{10}$Be are produced
as well as $^{8}$Li in Li convertor.
However,
$^{3}$H has a long half-life of $\sim$12.3 y.
Also, production yields of $^{6}$He and $^{10}$Be
are much lower compared to those of $^{8}$Li
by factors $\sim$10$^{4}$ and $\sim$10$^{7}$, respectively.
Other unstable isotopes,
$^{10}$Be, $^{12}$B and $^{14}$C,
are also produced in graphite materials.
Because of very low yields for $^{10}$Be and $^{12}$B
and a long half-life of $^{14}$C ($\sim$5.7 $\times$ 10$^{3}$ y),
their contributions are marginal for the IBD neutrino
detections.\footnote{\protect
For lead and tungsten as a reflector material,
$^{205, 209}$Pb and $^{181, 185, 187}$W isotopes
are produced.
But, $^{205}$Pb has a long half-life of $\sim$1.73 $\times$ 10$^{7}$ y, and
$^{181}$W emits low energy electron neutrinos (Q = 0.188 MeV), but does not electron antineutrinos.
And the ${\bar{\nu}}_{e}$ from
$^{209}$Pb (Q = 0.644 MeV), $^{185}$W (Q = 0.432 MeV) and $^{187}$W (Q = 1.31 MeV)
have the endpoint energies less than the IBD reaction threshold (1.806 MeV). Therefore, they do not affect the main IBD reaction.
}

%%%%%%%%%%%%%%%%%%%%%%%%%%%%%%%%%%%%%%%%%%%%%%%%%%%%
\begin{figure}[tbp]
\begin{center}
\epsfig{file=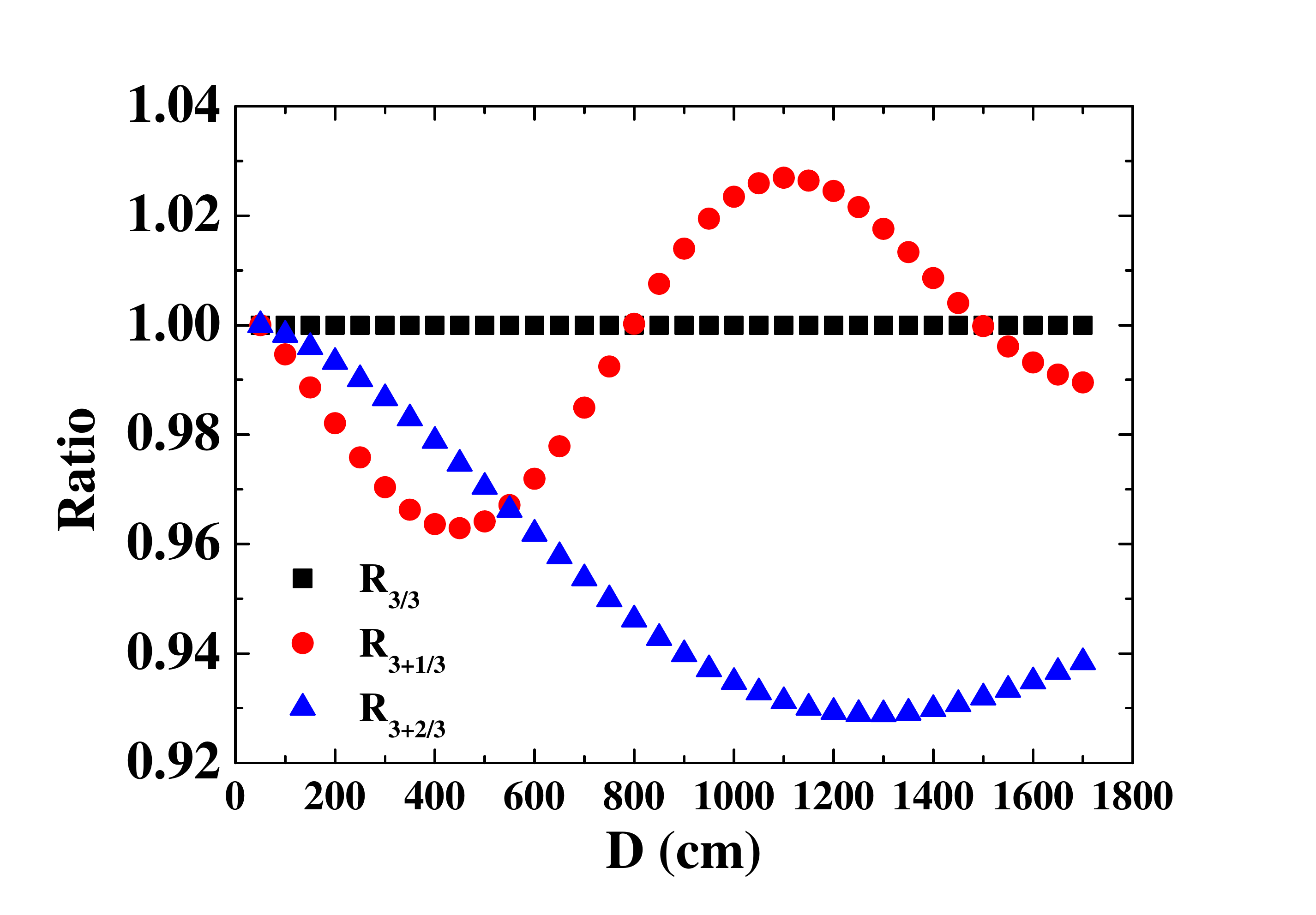, width=4.in}
\end{center}
\caption{(Color online)
The event ratios ER$_{3}$, ER$_{3+1}$ and ER$_{3+2}$ of ER$_{3}$
with respect to D.
}
\label{TotR}
\end{figure}
%%%%%%%%%%%%%%%%%%%%%%%%%%%%%%%%%%%%%%%%%%%%%%%%%

Background neutrinos such as
neutrinos from fission product of $^{252}$Cf ($\nu_{f}$) and
geo-neutrinos ($\nu_{geo.}$) can also affect the neutrino detection.
To check the $\nu_{f}$ effect,
we evaluate flux and event rate for $\nu_{f}$ by using
ENDF/B-VII.1 and ENSDF data.
Figure \ref{Fy_252Cf} shows the yields of fission products from
$^{252}$Cf isotope where the yield data of 1245 isotopes in the region with
30 $<$ Z $<$ 70 and 40 $<$ N $<$ 110 are taken from ENDF/B-VII.1.

First, we estimate the $\nu_{f}$ flux by using ENSDF data.
In our estimation,
we assume that unstable isotopes (880 isotopes) are fully decay with half-lifes shorter than 1 yr.
Figure \ref{figFISS} shows
electron antineutrino flux from $^{252}$Cf
and the flux from $^{8}$Li.
In the figure,
the $\nu_{f}$ is dominant
in the low energy region below the IBD reaction threshold
compared with the flux from $^{8}$Li.
However,
in the energy region of E$_{{\bar{\nu}}_{e}} >$ 5.1 MeV,
the neutrino flux from $^{8}$Li becomes larger than $\nu_{f}$ flux.
It is also found that
total event ratios by the neutrino from $^{8}$Li ($\nu_{^{8}Li}$)
for E$_{{\bar{\nu}}_{e}} >$ 7 MeV
is larger than those from $\nu_{f}$
by three orders of magnitudes
and thus contributions of the $\nu_{f}$ are negligible compare to
the $\nu_{^{8}Li}$.
Consequently, we can remove the $\nu_{f}$ effect with the neutrino energy cut of 7 MeV
in this work.

%geo neutrino
Geo-neutrinos, which can affect the neutrino detectors,
are produced via $\beta$-decays of long-lived radioactive isotopes
such as $^{40}$K, $^{238}$U and $^{232}$Th
that are present in the Earth.
$^{40}$K isotopes decay into $^{40}$Ca isotopes (branching ratio = 89.14\%)
through $\beta^{-}$-decays with the Q value of 1.311 MeV
and thus emit electron antineutrinos.
However, the neutrinos, which have the maximum energy near to the Q value, cannot affect the IBD reaction
because of a kinematic threshold of 1.806 MeV for the IBD.
Numbers of unstable isotopes are generated in
the chains of $^{238}$U and $^{232}$Th.
KamLAND \cite{geoKam_0} and Borexino \cite{geoBorex_0, geoBorex_1} have measured
a rate for $\nu_{geo.}$ ($\sim$ a few events/(100 ton $\cdot$ yr))
due to the decay of U or Th in the Earth.
The end points of the neutrino energy spectrum
from $^{232}$Th and $^{238}$U chain
are about 2.25 MeV and 3.3 MeV, respectively.
In this work,
we use the neutrino energy cut of 7 MeV.
Therefore,
contributions of the $\nu_{geo.}$ are also negligible compared to those
of ${\bar{\nu}}_{e}$ from $^{8}$Li ($\nu_{^{8}Li}$).

%%%%%%%%%%%%%%%%%%%%%%%%%%%%%%%%%%%%%%%%%%%%%%%%%%%
\begin{figure}[tbp]
%\begin{center}
\centering
\includegraphics[scale=0.5]{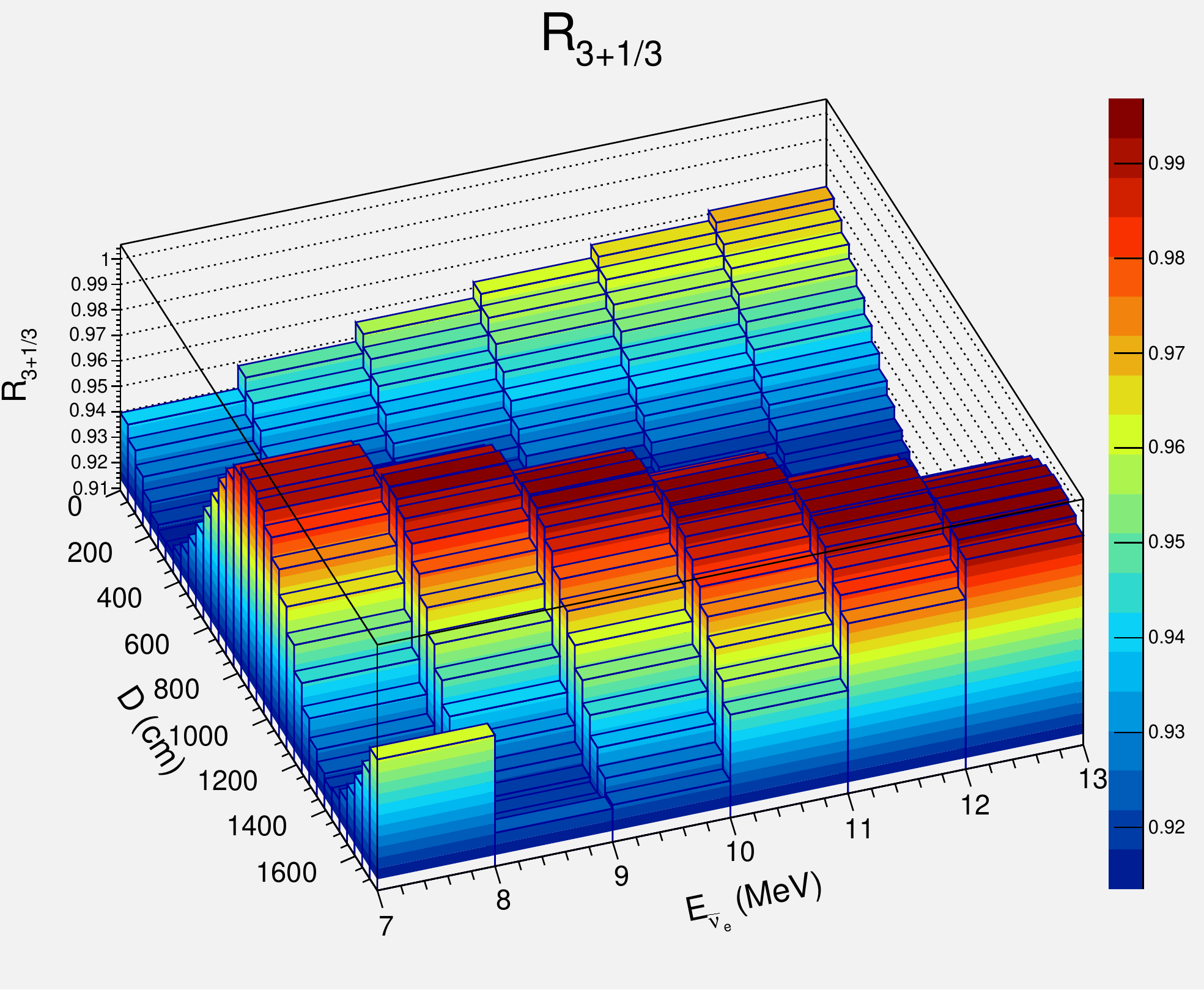}
\includegraphics[scale=0.5]{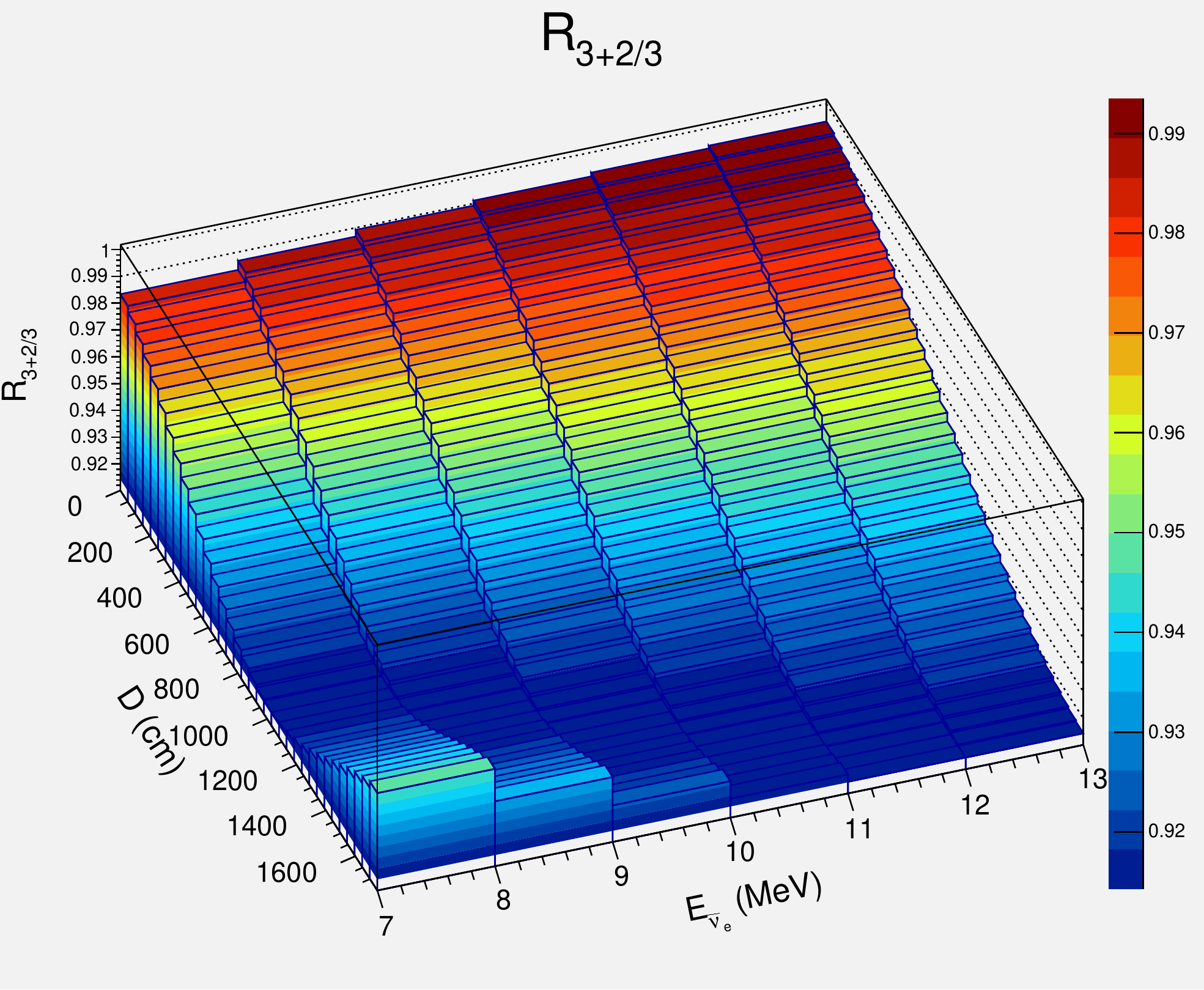}
%\end{center}
\caption{(Color online)
${\bar{\nu}}_{e}$ energy distribution of R$_{3+1/3}$ and R$_{3+2/3}$
with respect to D.
}
\label{RcomFig}
\end{figure}
%%%%%%%%%%%%%%%%%%%%%%%%%%%%%%%%%%%%%%%%%%%%%%%%%

Other radiations such as neutrons and gammas
can affect detection for neutrinos.
But, by
using a $^{252}$Cf source with a low intensity
and small detectors (for neutrons and gammas detections),
we can measure the background radiations from the setup suggested in this work.

\subsection{Spectral shape analysis for short baseline electron antineutrino disappearance studies
\label{sec4}}

%%%%%%%%%%%%%%%%%%%%%%%%%%%%%%%%%%%%%%%%%%%%%%%%%%%
%\begin{figure}[tbp]
%%\begin{center}
%\centering
%\includegraphics[scale=0.5]{R3to1_2.pdf}
%\includegraphics[scale=0.5]{R3to2_2.pdf}
%%\end{center}
%\caption{(Color online)
%${\bar{\nu}}_{e}$ energy distribution of R$_{3+1/3}$ and R$_{3+2/3}$
%with respect to D.
%}
%%
%\label{RcomFig2}
%\end{figure}
%%%%%%%%%%%%%%%%%%%%%%%%%%%%%%%%%%%%%%%%%%%%%%%%%

In order to see the effect by the sterile neutrinos,
we consider $^{8}$Li generator and hemisphere shape LS detector based on the JUNO \cite{JunoR},
as explained in Fig. \ref{figG_IBD}.
The expected event rates under P$_{3}$ model are compared with
the P$_{3+1}$ and the P$_{3+2}$ models,
and distinct features for each model are obtained.
Figure \ref{TotR} shows results for
the expected total event ratios (ERs) of P$_{3}$, P$_{3+1}$ and P$_{3+2}$ to P$_{3}$
as a function of D, which is the distance between bottom of the plate and center
of the shell for detection in Fig. \ref{figG_IBD}.
R$_{3/3}$ (R$_{3+1/3}$ and R$_{3+2/3}$) means
the ratio ER$_{3}$ (ER$_{3+1}$ and ER$_{3+2}$) to ER$_{3}$.
If there are no sterile neutrinos, the ratio should be 1 in Fig. \ref{TotR}.

Under the 3+1 sterile neutrino scenario with the best fit parameter, however,
oscillation shape is largely deviated from the unit number 1. The minimum and the maximum are shown
at D = 450 cm and D = 1100 cm, respectively.
For the 3+2 scenario,
the ratio decreases for D $<$ 1200 cm with the D increase, and
the period of the oscillation is found to be much longer than
that of the 3+1 scenario.
At D = 1100 cm,
the minimum for the 3+2 scenario
and the 3+1 scenario maximum appear clearly.
The neutrino energy distributions for different D values with dD = 1 m
are shown in Fig. \ref{RcomFig}.
It is found that
the energy oscillation shapes are also varying as the D changes.

%The visible energy (E$_{vis}$) of the prompt signal
%due to a
%positron ($e^{+}$)
%is strongly correlated with
%the energy of ${\bar{\nu}}_{e}$ (E$_{{\bar{\nu}}_{e}}$),
%E$_{{\bar{\nu}}_{e}}$ $\simeq$ E$_{vis}$ + 0.78 MeV.
%${\bar{\nu}}_{e}$ energy spectra can be reconstructed
%using the E$_{vis}$.

\begin{figure}[tbp]
\begin{center}
\epsfig{file=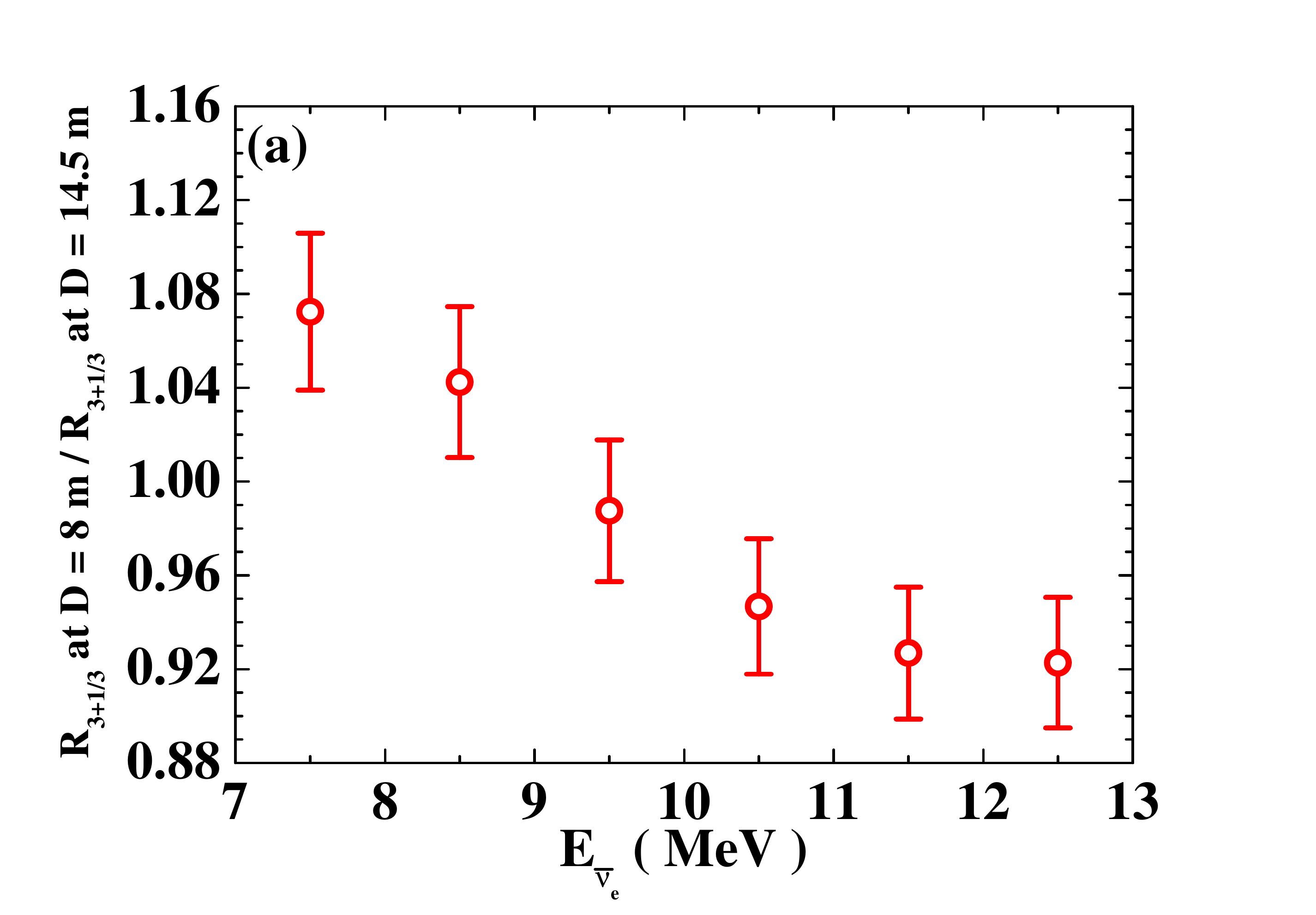, width=3.in}
\epsfig{file=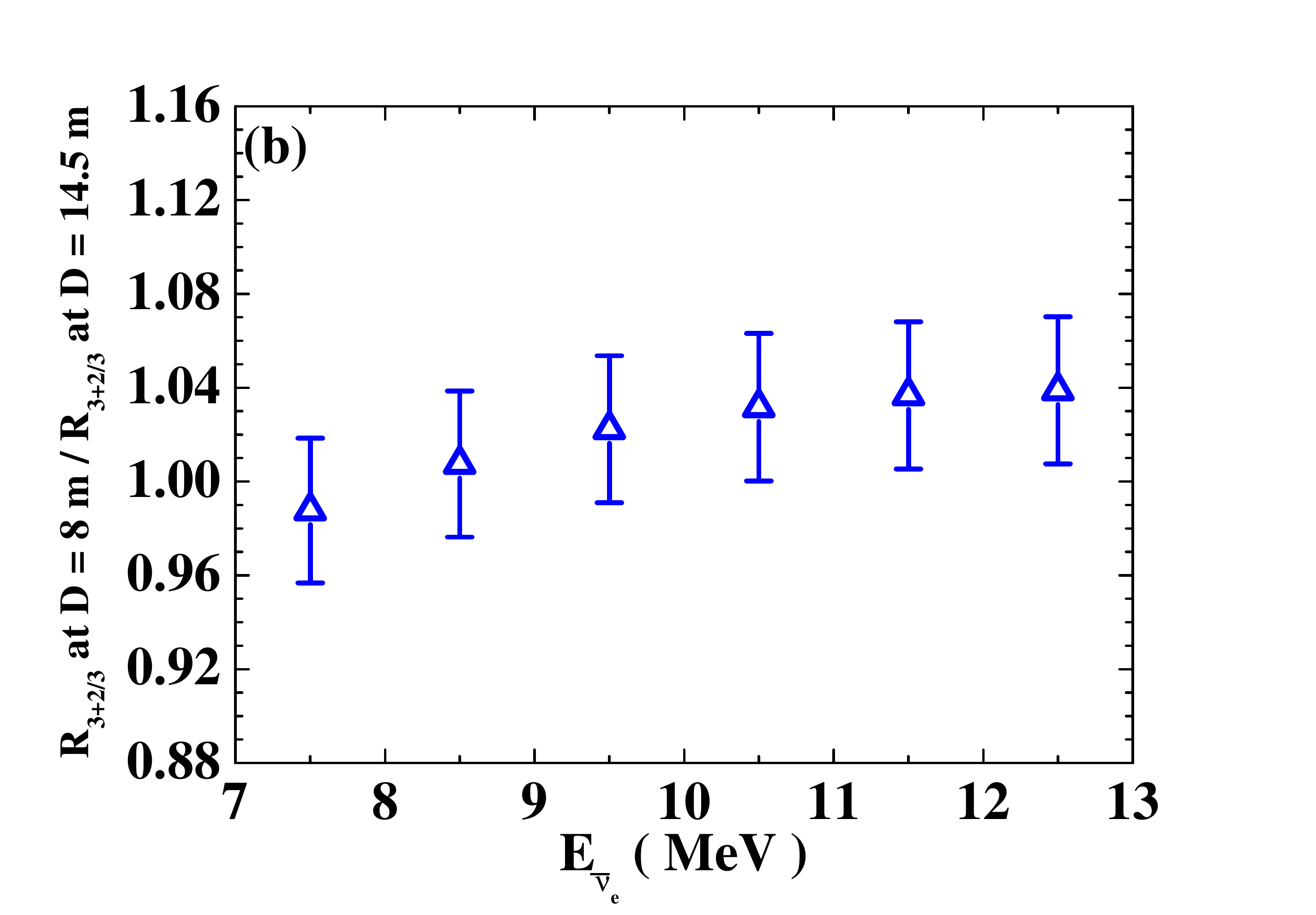, width=3.in}
\end{center}
\caption{(Color online)
Panel (a) is the expected ratio of R$_{3+1/3}$ at D = 8 m and D = 14.5 m with respect to E$_{{\bar{\nu}}_{e}}$. Panel (b) is for R$_{3+2/3}$ case.
}
\label{ComERatio}
\end{figure}

For the R$_{3+1/3}$ ratio in the figure,
the peak positions of neutrino energies
increase as the D increase.
On the contrary,
monotonic changes in the shapes
show up for R$_{3+2/3}$ ratio in our setup.
More drastic changes
for the 3+1 sterile neutrino scenario
can be obtained
from two distinct signals at two different detector positions, at D = 8 m and D = 14.5 m. For D = 8 m,
R$_{3+1/3}$ ratios are 0.917 and 0.988 at
E$_{{\bar{\nu}}_{e}}$ = 12.5 MeV and E$_{{\bar{\nu}}_{e}}$ = 7.5 MeV, respectively.
However, for D = 14.5 m, the situation is reversed. That is, R$_{3+1/3}$ ratios of 0.994 and 0.922 are obtained at
E$_{{\bar{\nu}}_{e}}$ = 12.5 MeV and E$_{{\bar{\nu}}_{e}}$ = 7.5 MeV, respectively.
Therefore, using the expected two different signals for
D = 8 m and D = 14.5 m,
we can easily search for 3+1 sterile neutrino scenario.
Figure \ref{ComERatio} shows the expected ratio of R$_{3+1/3}$ at D = 8 m and D = 14.5 m in panel (a), and that of R$_{3+2/3}$ is shown in panel (b).

%%%%%%%%%%%%%%%%%%%%%%%%%%%%%%%%%%%%%%%%%%%%%%%%%%%%
%%%%%%%%%%%%%%%%%%%%%%%%%%%%%%%%%%%%%%%%%%%%%%%%%
%

For the error bars in the analysis,
we assumed
a statistical error of 2\%,
a systematic error of 2\%,
an energy resolution of 3{\%}/$\sqrt{\rm{E/MeV}}$,
a position resolution of 12 cm,
and a IBD cross section error of 0.5\%.
Decrease pattern can be seen clearly in the Fig. \ref{ComERatio} (a)
as the E$_{{\bar{\nu}}_{e}}$ increase.
Also,
the maximum and the minimum values of the ratios
are 1.07 at E$_{{\bar{\nu}}_{e}}$ = 7.5 MeV and
0.92 at E$_{{\bar{\nu}}_{e}}$ = 12.5 MeV,
respectively.
If we can measure
approximately 15\% deviation from the
expected events,
we can find a clue to the problem of
whether the P$_{3+1}$
model is the most appropriate scenario.

\section{Summary
\label{sum-sec}}

In this work,
we propose an experimental setup for neutrino spectral shape analysis
by using a $^{8}$Li generator under non-accelerator system.
Unstable isotope, $^{8}$Li,
emits ${\bar{\nu}}_{e}$ having energy range
0 $<$ E$_{{\bar{\nu}}_{e}}$ $<$ 13 MeV ($<$~E$_{{\bar{\nu}}_{e}}$~$>$ $\sim$ 7 MeV)
through $\beta^{-}$ decay where
the $^{8}$Li is produced via $^{7}$Li(n,$\gamma$)$^{8}$Li reaction
with an intense neutron emitter, $^{252}$Cf.
The $^{8}$Li generator suggested in our scheme
can be placed and applied to any neutrino detectors
such as Borexino, JUNO, KamLAND, LENA and SNO+,
because accelerator or reactor systems are not needed anymore.

Suggested $^{8}$Li generator is very compact, so that neutrino detectors can be placed within a few meters
from the neutrino source
with E$_{{\bar{\nu}}_{e}}$ $<$ 13 MeV.
Moreover neutrinos from the source can be so effectively controlled.
Consequently, background neutrinos can be exactly separated as mentioned in Sec. 3.2.
It means that our scheme could be a very efficient neutrino source
for the study of 1 eV mass scale sterile neutrino
as well as other neutrino oscillation studies.

The expected event rates with P$_{3}$, P$_{3+1}$ and P$_{3+2}$ models with best fit points,
and their ratios are presented for different detector distances.
Our results show that
neutrino disappearance features and possible reaction rate
are changed significantly by the sterile neutrino.
These distinct features
can give useful chances to search for
the existence of a sterile neutrino
as well as the test of
the 3+1 or 3+2 sterile neutrinos
scenarios.
In particular,
if we can confirm 15\% deviation in Fig. \ref{ComERatio},
we can conclude whether P$_{3+1}$ model is the most appropriate sterile neutrino model or not.

\section*{Acknowledgments}
The work of J. W. Shin is supported by
the National Research Foundation of Korea \ (Grant No. NRF-2015R1C1A1A01054083),
the work of M.-K. Cheoun is supported by
the National Research Foundation of Korea \ (Grant No. NRF-2015K2A9A1A06046598 and NRF-2017R1E1A1A01074023).

\bibliography{mybibfile}

\newpage

\end{document}